\documentclass[sigconf]{acmart}

\AtBeginDocument{%
  \providecommand\BibTeX{{%
    \normalfont B\kern-0.5em{\scshape i\kern-0.25em b}\kern-0.8em\TeX}}}

\copyrightyear{2020}
\acmYear{2020}
\setcopyright{none}
\acmConference[MSR '20]{17th International Conference on Mining Software Repositories}{October 5--6, 2020}{Seoul, Republic of Korea}
\acmBooktitle{17th International Conference on Mining Software Repositories (MSR '20), October 5--6, 2020, Seoul, Republic of Korea}
\acmPrice{15.00}
\acmDOI{10.1145/3379597.3387505}
\acmISBN{978-1-4503-7517-7/20/05}

\begin{document}
\begin{sloppy}

\title[Impact of the Dynamics of Collaboration on Introverts]{The Impact of Dynamics of Collaborative Software Engineering on Introverts: A Study Protocol}

\author{Ingrid Nunes}
\affiliation{%
  \institution{Universidade Federal do Rio Grande do Sul (UFRGS)\\Instituto de Inform\'{a}tica}
  \city{Porto Alegre}
  \country{Brazil}}
\email{ingridnunes@inf.ufrgs.br}

\author{Christoph Treude}
\affiliation{%
  \institution{University of Adelaide\\School of Computer Science} 
  \city{Adelaide}
  \country{Australia}}
\email{christoph.treude@adelaide.edu.au}

\author{Fabio Calefato}
\affiliation{%
  \institution{University of Bari\\Dipartimento di Informatica}
  \city{Bari}
  \country{Italy}
}
\email{fabio.calefato@uniba.it}


\begin{abstract}
\textbf{Background}: Collaboration among software engineers through face-to-face discussions in teams has been promoted since the adoption of agile methods. However, these discussions might demote the contribution of software engineers who are introverts, possibly leading to sub-optimal solutions and creating work environments that benefit extroverts. \textbf{Objective}: We aim to evaluate whether providing software engineers with time to work individually and reason about a collective problem is a setting that makes introverts more comfortable to interact and contribute more, ultimately leading to better solutions. \textbf{Method}: We plan to conduct a between-subjects study, with teams in a control group that design a software architecture in a team discussion meeting and teams in a treatment group in which subjects work individually before engaging in a meeting. We will assess and compare the amount of contribution of introverts, their subjective experiences, and the designed solutions. \textbf{Limitations}: As extroverts will be present in both groups, we will not be able to conclude that better solutions are solely due to the increased participation of introverts. The analyses of their subjective experience and amount of contributions might provide evidence to suggest the reasons for observed differences.
\end{abstract}



\maketitle

\section{Introduction}

In traditional software management~\cite{Fairley:book2009:ProjectMgnt}, a project manager typically assigns tasks to software engineers who then work individually on these tasks. With the wide adoption of agile methods~\cite{Martin:book2003:Agile}, the way of developing software has shifted to a more collaborative environment. Software engineers are organized in self-managing teams, in which involved engineers have daily meetings to synchronize their work, may work in pairs, have face-to-face discussions, and have retrospective meetings to adapt their dynamics. These practices promote a significant amount of \emph{interaction} among individuals, which is assumed to lead to better solutions and welcoming environments. This change in the dynamics within software teams is complemented by changes in the physical work environment that are now commonly configured as open workspaces.

Promoting interaction, mainly in person, in software engineering has a large impact on individuals. Humans have different characteristics, including personality traits. Thus, adopted practices may be experienced differently by each individual. In her book, \citeauthor{Cain:book2013:Introversion}~\cite{Cain:book2013:Introversion} discussed the ``\emph{extroverted ideal},'' which is the current trend to consider extroversion as a desired personal characteristic~\cite{Paulhus:1997} and expect that those who are introverted behave as extroverts.
As collaborative environments require interaction, they may be more adequate for those who are extroverted, causing introverts to contribute less, because small signs of disapproval~\cite{Cain:book2013:Introversion} may cause them not to share their ideas. Therefore, as argued by \citeauthor{Cain:book2013:Introversion}, there are occasions in which ``collaboration kills creativity''~\cite{Dunnette:1963}. 

In response, in this registered report, we detail a study protocol in which we investigate work practices used to produce a collective software solution and how they are experienced by extrovert and introvert software engineers. In a nutshell, the study requires subjects to jointly design a software architecture to a given software problem. Teams in the control group have a fixed time to discuss in a face-to-face meeting and propose a solution. Teams in the treatment group have this same fixed time, but in the first half of the time, they work individually on the problem. We compare the outcomes of the teams and how extroverts and introverts contributed to the solution. We also assess their subjective experiences. 

\section{Background and Related Work}~\label{section:relatedWork}

\paragraph{Personality Trait Theories and Big Five Model}
Personality is the set of all behavioral, emotional, and mental attributes that characterize a unique individual~\cite{mairesse2007}. Psychologists have sought for years to formulate descriptive models of high-level traits that would provide a framework to simplify the organization and description of the major individual differences among human beings~\cite{john1999}.
Many personality trait theories have been proposed since the 1930s, although more general acceptance and interest was not achieved until the 1970s when research began to find empirical evidence on the validity of a general taxonomy of five orthogonal personality traits now referred to as the \textit{Big Five Model}~\cite{goldberg1981} or \textit{Five Factor Model}~\cite{mccrae1987}.

According to the Big Five Model, the most important individual characteristics can be described by the following five orthogonal dimensions (often referred to using the OCEAN mnemonic).

\begin{itemize}
  \item \relax \textbf{\textit{Openness}} (inventive/curious vs.\ consistent/cautious): the extent to which a person is open to  experiences; people low in Openness tend to be more conservative and close-minded.
  \item \relax \textbf{\textit{Conscientiousness}} (efficient/organized vs.\ easy-going/ careless): the tendency to plan in advance in goal-directed behavior; low-Conscientiousness individuals are more tolerant and less bound by rules and plans.
  \item \relax \textbf{\textit{Extroversion}} (outgoing/energetic vs.\ solitary/introverted): the tendency to seek stimulation in the company of others; introverted individuals who are low in Extroversion are reserved and solitary.
  \item \relax \textbf{\textit{Agreeableness}} (friendly/compassionate vs.\ challenging/detached): the tendency to be compassionate and cooperative toward others; low Agreeableness is related to being suspicious, challenging, and antagonistic.
  \item \relax \textbf{\textit{Neuroticism}}\textit{} (sensitive/nervous vs.\ secure/confident): the emotional stability is the extent to which a person's emotions are sensitive to the environment; those who have a low score in Neuroticism are calmer and more stable, while neurotic individuals are prone to psychological distress and anxiety.
\end{itemize}

\paragraph{Personality in Software Engineering}
The study of personality in software engineering has drawn the attention of researchers for decades. In the early 1970s, \citeauthor{weinberg1971}~\cite{weinberg1971} was the first to hypothesize that personality could impact the performance of software engineers. Later, in the 1980s, \citeauthor{shneiderman1980}~\cite{shneiderman1980} argued that the personality of software engineers could play a critical role in determining how they interact.
Since then, a growing amount of research has been conducted on the effects of personality in software engineering. In their systematic literature review, \citeauthor{cruz2015-slr}~\cite{cruz2015-slr} identified 90 studies conducted between 1979 and 2014, most of which (about 70\%) were published after 2002.
Previous studies on personality in software engineering have focused on different aspects, such as the prediction of performance~\cite{hannay2010,karimi2016}, work preferences~\cite{kosti2014,raza2014}, job satisfaction~\cite{acuna2009,acuna2015}, and team composition~\cite{dasilva2013,gilal2016}.

\paragraph{Team Composition in Software Engineering}
Software team composition has also been studied  from perspectives other than personality. \citeauthor{siau2010characteristics}~\cite{siau2010characteristics} interviewed 21 professional software engineers and used open coding to derive a list of fifty-nine unique characteristics, classified into eight categories. Among these categories, attitude/motivation, knowledge, interpersonal/communication skills, and working/cognitive ability were perceived by the interviewees as the most important.
\citeauthor{kang2006-smm}~\cite{kang2006-smm} investigated the importance of team member characteristics, particularly cognitive and demographic, on software team effectiveness. They found that cognitive similarities, modeled via the construct of a Shared Mental Model, have a stronger influence than age, tenure, and gender similarity.
\citeauthor{wickramasinghe2015diversity}~\cite{wickramasinghe2015diversity} investigated the effects of diversity in global software team composition. By interviewing 216 software engineers involved in global software projects, they found that diversity is associated with conflicts within teams. However, they also found that when such conflicts are resolved with team leader support, diversity is helpful in achieving higher levels of team performance.


\paragraph{Work Practices and Work Spaces in Software Engineering}

In addition to personality and team composition, work practices of software engineers potentially impact (perceived) productivity and well-being. \citeauthor{meyer2019today}~\cite{meyer2019today} characterized the daily life of software engineers and found that they value being in control of their own workday, without disruptions by external factors or deviations from plans. Collaboration plays a central role in the daily activities of software engineers, with estimates as high as 45\% of work time~\cite{gonccalves2011collaboration}.

How this collaboration takes place influences its outcome. While \citeauthor{bird2009does}~\cite{bird2009does} found only negligible differences in failure rates between components developed in distributed settings and components developed in collocated settings, \citeauthor{damian2007collaboration}~\cite{damian2007collaboration} report that distance affects how accessible remote colleagues are. Even when collocated, work spaces can be set up differently: \citeauthor{mishra2012impact}~\cite{mishra2012impact} found that half cubicles are very effective for the frequency of communication and that half-height glass barriers are very effective during individuals' problem-solving activities while working together as a team. The authors conclude that such a physically open environment appears to improve communication, coordination, and collaboration. In a remote setting, \citeauthor{damian2008need}~\cite{damian2008need} compared teams of stakeholders using synchronous videoconferencing for requirements negotiations to teams with an additional asynchronous text-based discussion phase before the video conference and found that teams with initial asynchronous discussions were more effective. Our proposed study follows a somewhat similar design.

\section{Research Question and Hypotheses}

Our goal with this study is to promote better work environments considering the personal traits of individuals, possibly leading to better software solutions. In particular, we focus on the level of extroversion. In this context, we state the following research question: \emph{Do introvert software engineers contribute more and feel more comfortable to contribute when they are given time to work individually?} Considering this research question, there are three hypotheses that we aim to test, described as follows.

\begin{description}
	\item[H1.] Introverted software engineers prefer to work individually before engaging in a team discussion to produce a software solution.
	\item[H2.] When introverted software engineers are given time to work individually before engaging in a team discussion to produce a software solution, they contribute more than if no time for individual work is given. 
	\item[H3.] A team of software engineers produce better software solutions if team members are given time to work individually before engaging in a team discussion than if no time for individual work is given.
\end{description}

Our study focuses on understanding the behavior of introverted software engineers. However, considering the existing literature on introversion, the presence of extroverts in team discussions may be a factor that discourages introverts to participate actively. Therefore, as discussed later, our study involves teams of both introverts and extroverts. Consequently, H3 is related to the outcome produced by each group, which includes extroverts. H1 and H2 help to understand the causes of differences that are possibly identified when testing H3.

\section{Variables}\label{sec:variables}

There are two independent variables associated with our three hypotheses. The first is the \textbf{personality trait} of each subject, which can be \emph{extroverted} or \emph{introverted}. This is obtained by means of answers to questions of the pre-questionnaire filled in by subjects. The second refers to our intervention, which is the \textbf{interaction model}. There are two alternatives: (i) \emph{team only}, when there is a single team discussion occurring in two time slots; and (ii) \emph{individual and team} when in the first time slot team members work individually and in the second time slot there is a discussion.

The dependent variables are different for each hypothesis. For H1, we assess the answers to questions of the post-questionnaire provided by each subject on a 7-point Likert scale. For H2, we measure the percentage of the total time of team discussion in which the subject speaks up. For H3, we collect from experts in software engineering scores from 0 to 10 that indicate the quality of the solution provided by each team.

Finally, a possible confounding variable that we control is the \emph{expertise} of subjects. The pre-questionnaire that subjects are required to answer includes questions to assess their expertise in software engineering.

\section{Material and Tasks}

In this section, we describe pre- and post-questionnaires, the software problem, and the expert evaluation for the proposed study. The study materials are available online.\footnote{\url{https://www.inf.ufrgs.br/prosoft/resources/2020/msr-rr-introversion/}}

\subsection{Pre-questionnaire}

Subjects are required to answer a pre-questionnaire, composed of four parts, before participating in the study. In the first part, they are asked to sign the informed consent and agreement to have their data processed. In the second part, they inform us of their demographic characteristics: age, gender, nationality, and education. Then, in the third part, we request them to share their experience in software engineering, detailing their current position, years of experience, professional experience, and self-reported knowledge on software architecture, modularity, architectural patterns, design patterns, web development, and programming. Finally, they answer the questions of the IPIP test\footnote{\url{https://ipip.ori.org/}} associated with extroversion/introversion. 

\subsection{Software Problem}

In our study, subjects are required to propose, in teams, a solution to a software problem. This problem is given as a software system to be developed, involving functional and non-functional requirements (such as security, scalability, and reliability). Teams are informed that they need to design a software architecture that satisfies these requirements and also follows principles of software engineering. The architecture should be described in terms of modules, their roles, dependencies among them, the control flow to process requests, and used technologies. Each team provides a single solution to the problem. The team discussions are recorded so that we can collect the amount of time during which each subject speaks up.

\subsection{Post-questionnaire}

After performing the key task of our study, i.e., the proposal of an architecture for a software system, subjects are requested to answer a post-questionnaire reporting their experience. Answers are given on a 7-point Likert scale. We ask them if they: (i) felt comfortable to share their ideas; (ii) felt respected by their team members; (iii) felt confident to contribute; (iv) enjoyed participating in the team discussion; (v) (would have) enjoyed having time to work individually; and (vi) had a positive experience while performing the requested task. Subjects can also share additional comments.

\subsection{Expert Evaluation}

To evaluate the quality of the solutions provided by subjects, we ask three experts in software engineering, in particular software architecture, to inspect all solutions. They blindly provide a score ranging from 0 (worst) to 10 (best) for each of the following aspects: overall evaluation, functional requirements, non-functional requirements, understandability, and modularity. The scores must be justified. After providing an initial score (first round), if there is no convergence, we compile all scores and justifications given by experts in a single document with anonymized data and return them to the experts, who must reassess their scores (second round). We perform at most five rounds to achieve convergence. If this is not the case, they have a meeting to discuss and converge (in this case, they will become aware of their identity). This evaluation method is inspired by the Wideband delphi estimation method~\cite{valerdi201110}.

\section{Subjects}

We select candidates to participate in the study using convenience sampling. All involved researchers reach out to professional software engineers with whom they have contact and request for volunteers to participate in the study. All volunteers are requested to complete the pre-questionnaire. From these, we select only extroverts and introverts, excluding extroverts with introvert tendencies and introverts with extrovert tendencies. This will be done considering participants with high and low scores, based on the absolute score or distribution of scores in our sample.\footnote{\url{https://ipip.ori.org/InterpretingIndividualIPIPScaleScores.htm}} This choice depends on the scores of our candidates. We also exclude volunteers that do not have a degree in Computer Science (or similar courses), do not have professional experience, or do not have at least average knowledge on the topics listed in the pre-questionnaire. Our goal is to have 16 teams of 4 (assigned randomly) subjects each (2 extroverts and 2 introverts). Subjects of the same team must be in the same geographical location. If more candidates are eligible than our goal, we select subjects randomly.

\section{Execution Plan}

Our study follows a between-subjects design. It is composed of the following steps.

\begin{enumerate}
	\item After selecting the candidates for the study, we request these candidates to complete the pre-questionnaire. 
	\item We analyze the answers to the pre-questionnaire and select our sample as described in the previous section.
	\item Randomly, we select half of the formed teams to receive the intervention. 
	\item Each team must provide a solution to the given problem. Teams in the control group have two time slots (90~min total) to have a discussion with all members and propose a solution. Teams in the treatment group have one time slot (45~min) for members to work individually towards a solution and one time slot (45~min) to discuss with all members to propose a solution. All discussion meetings are recorded.
	\item Subjects are requested to complete the post-questionnaire.
	\item We ask three experts in software engineering to evaluate the proposed solutions according to the described \emph{Expert Evaluation} method.
\end{enumerate}

Based on the collected data, we extract the values of the variables detailed in Section~\ref{sec:variables} for analysis.

\section{Analysis Plan}

\noindent\textbf{H1 and H2.} The results are described detailing the minimum, maximum, average, and median values. The answers provided to the post-questionnaire are also presented in Likert plots, while active participation in the team discussions is presented in box plots. To test for significant differences between the four groups (introvert-control, introvert-treatment, extrovert-control, extrovert-treatment), we use a two-way ANOVA test, if its assumptions are met. If not, we use a Kruskal-Wallis test. A corresponding \textit{post-hoc} test is used if the test shows a significant difference. 

\noindent\textbf{H3.} As H3 tests for differences between the solutions provided by teams in the control group and teams in the treatment group, there are only two groups to compare. We test for differences between scores given to each evaluated aspect of the provided solutions. We first test for normality, using a Shapiro-Wilk test. If the distribution is normal, we use a parametric test (t-test); otherwise, we use a non-parametric test (Wilcoxon). Descriptive statistics are given similar to the ones given for the dependent variables associated with H2.


\bibliographystyle{ACM-Reference-Format}
\bibliography{references}

\end{sloppy}
\end{document}